\documentclass{moriond}

\def\Journal#1#2#3#4{{#1} {\bf #2}, #3 (#4)}

\def\NPB{{\em Nucl. Phys.} B}
\def\PLB{{\em Phys. Lett.}  B}
\def\PRL{\em Phys. Rev. Lett.}
\def\PRD{{\em Phys. Rev.} D}

\def\be{\begin{equation}}
\def\ee{\end{equation}}
\def\bea{\begin{eqnarray}}
\def\eea{\end{eqnarray}}

\usepackage{amssymb}

\newcommand{\Photo}{\includegraphics[height=35mm]{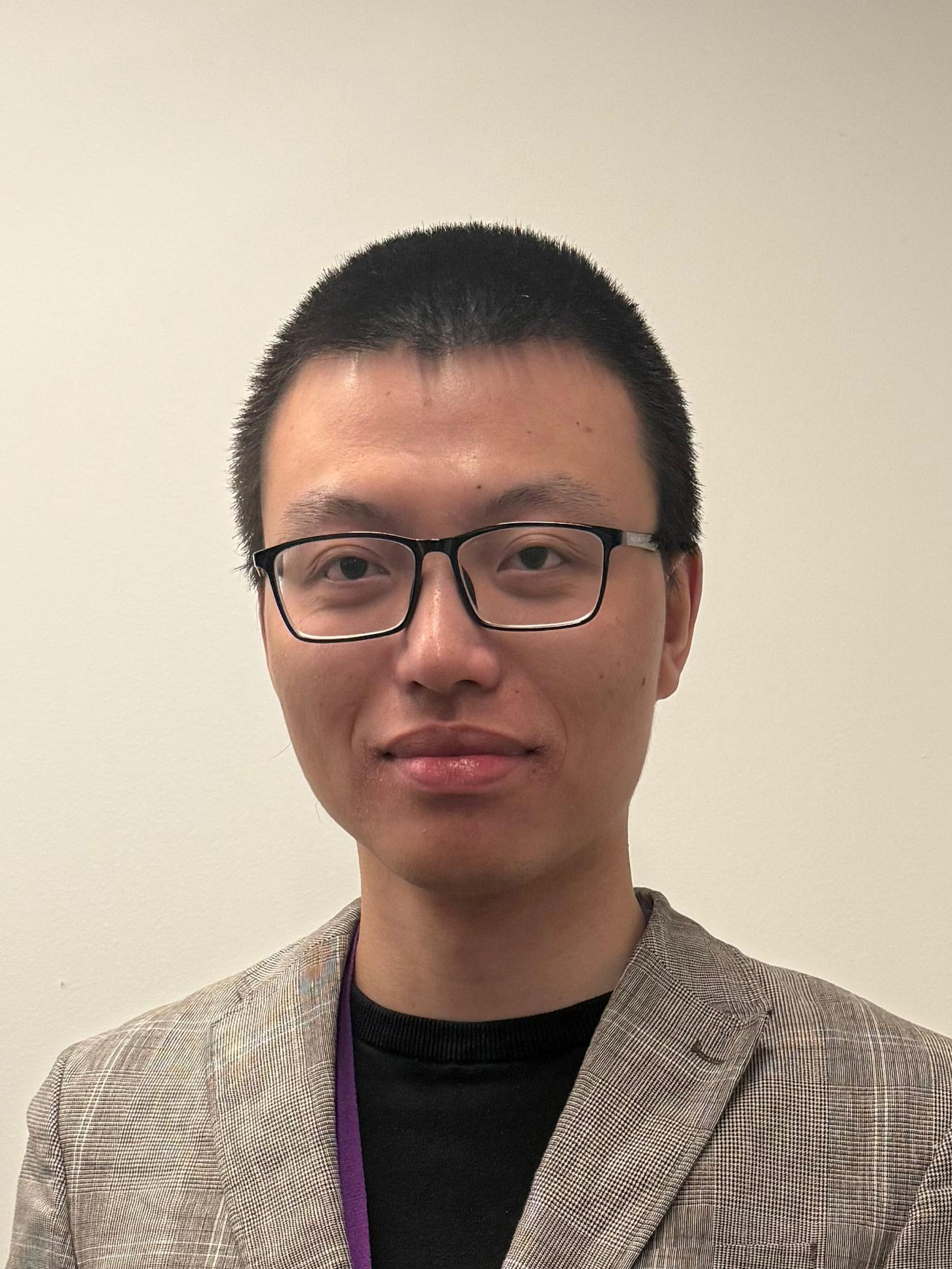}}

\begin{document}
\vspace*{4cm}
\title{New measurement of $K^+\to\pi^+\nu\bar\nu$ branching ratio at the NA62 experiment}

\author{
Xiafei Chang~\footnote{for the NA62 Collaboration:
A.~Akmete, R.~Aliberti, S.~Alibocus, F.~Ambrosino, R.~Ammendola,  A.~Antonelli, G.~Anzivino, R.~Arcidiacono,
A.~Baeva, D.~Baigarashev
% ...
}
}

\address{Laboratoire de Physique des Hautes Energies, EPFL,\\
1015 Lausanne, Switzerland}

\maketitle\abstracts{
The ultra-rare decay $K^+\to\pi^+\nu\bar\nu$ is a golden mode in flavor physics. The Standard Model prediction for its branching ratio is below $10^{-10}$. This decay mode is highly sensitive to new physics models at mass scales up to $\mathcal{O}(100\,\mathrm{TeV})$. The NA62 experiment at CERN SPS is designed to measure this decay mode. A preliminary result of the branching ratio measurement using data collected in 2023--2024 is presented. With the new dataset, the NA62 experiment doubled its signal sample while reducing the background in proportion. Combining the data collected in 2016--2024, the branching ratio is measured to be $\mathcal{B}(K^+\to\pi^+\nu\bar\nu) = \left(9.6^{+1.9}_{-1.8}\right)\times10^{-11}$. The result is compatible with the Standard Model prediction with a precision better than $20\,\%$.
}

\section{Introduction}

The decay $K^+\to\pi^+\nu\bar\nu$ is a Flavor Changing Neutral Current (FCNC) process. This decay proceeds at the lowest order in the Standard Model (SM) through electroweak box and penguin diagrams and is therefore highly suppressed by the Glashow-Iliopoulos-Maiani
(GIM) mechanism and the Cabibbo-Kobayashi-Maskawa (CKM) matrix element governing $t\to d$ quark transitions. Since the decay is dominated by the short-distance $t$-quark exchange and the hadronic matrix element can be extracted from the precisely measured decay of $K^+\to\pi^0e^+\nu$~\cite{Brod:2010hi,Isidori:2005xm,Mescia:2007kn}, the SM predicts the branching ratio of $K^+\to\pi^+\nu\bar\nu$ with a precision better than $8\,\%$~\cite{Buras:2022wpw,KaonsAtCERN2023,Allwicher:2024ncl}. The uncertainty is dominated by the CKM parameters $V_{cb}$ and $\gamma$, while the intrinsic theoretical uncertainty is approximately $3\,\%$~\cite{Brod:2010hi}. The prediction of the branching ratio was $\mathcal{B}(K^+\to\pi^+\nu\bar\nu)=(8.4\pm1.0)\times10^{-11}$, using CKM matrix elements extracted from tree-level decays \footnote{Used as benchmark value in this analysis to compare with the previous results.}~\cite{Buras:2015qea}. More recent works predict $(8.60\pm0.42)\times10^{-11}$, using only meson mixing to eliminate $|V_{cb}|$ dependence~\cite{Buras:2022wpw}, and $(7.86\pm0.61)\times10^{-11}$, using a full CKM parameter fit~\cite{DAmbrosio:2022kvb}.

The $K^+\to\pi^+\nu\bar\nu$ decay is among the most promising channels to search for physics beyond the Standard Model (BSM). It probes new physics at mass scales up to $\mathcal{O}(100\,\mathrm{TeV})$. Several BSM scenarios predict significant deviations from the SM branching ratio, correlations with other flavor observables, as well as correlations with the corresponding neutral mode $K_L\to\pi^0\nu\bar\nu$~\cite{Buras:2015yca,Deppisch:2020oyx,Crosas:2022quq,Buras:2024ewl}. A model-independent relationship between the charged and neutral modes is provided by the Grossman-Nir bound~\cite{Grossman:1997sk}: $\mathcal{B}(K_L\to\pi^0\nu\bar\nu)\lesssim 4.3\cdot\mathcal{B}(K^+\to\pi^+\nu\bar\nu)$. The direct measurement of the neutral mode is conducted by the KOTO experiment at J-PARC, providing the latest upper limit $\mathcal{B}(K_L\to\pi^0\nu\bar\nu)<2.2\times10^{-9}$ at $90\,\%$ CL~\cite{KOTO:2024zbl}.

The first $K^+\to\pi^+\nu\bar\nu$ candidate event was reported by the E787 and E949 experiments at BNL, using the kaon decay-at-rest technique, with a measured~\cite{BNL-E949:2009dza} $\mathcal{B}(K^+\to\pi^+\nu\bar\nu)=17.3^{+11.5}_{-10.5}\times10^{-11}$. The NA62 experiment at CERN was designed to measure the $K^+\to\pi^+\nu\bar\nu$ decay with a kaon decay-in-flight technique. It provided evidence at the $3.4\sigma$ level using its 2016--2018 dataset and has achieved $5\sigma$ observation with the combined 2016--2022 dataset~\cite{PnnRun1Paper,Pnn2122paper}. The measured branching ratio was $\mathcal{B}(K^+\to\pi^+\nu\bar\nu)=13.0^{+3.3}_{-3.0}\times10^{-11}$, from an observation of $51$ events with $18^{+3}_{-2}$ expected background events. The NA62 experiment continues to collect data until 2026. The preliminary result of the $K^+\to\pi^+\nu\bar\nu$ measurement with the dataset collected by the NA62 experiment in 2023--2024 are reported below.

\section{The NA62 experiment}\label{sec:na62exp}

\begin{figure}
    \centering
    \includegraphics[width=\linewidth]{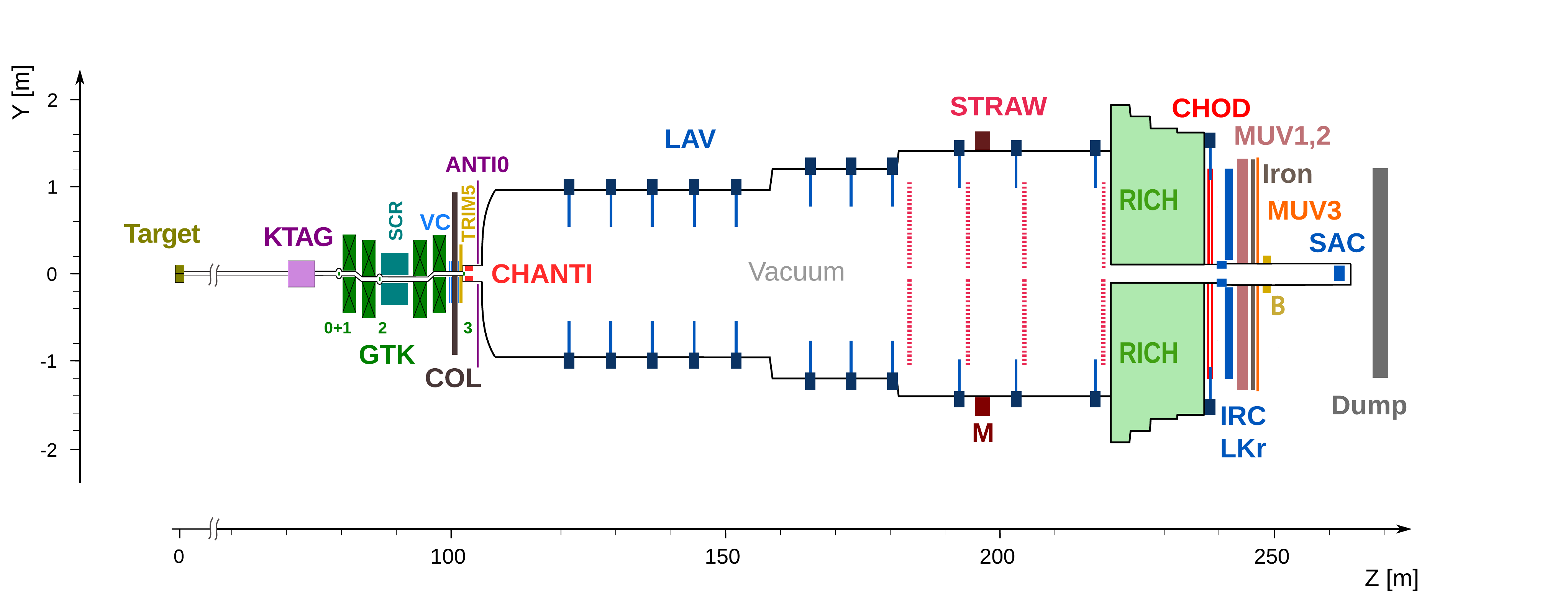}
    \caption[The NA62 detector]{Schematic side view of the NA62 detector for data-taking from 2021 onwards.}
    \label{fig:na62exp}
\end{figure}

A detailed description of the NA62 beam line and detector can be found in Ref.~\cite{NA62DetectorPaper}. Fig. \ref{fig:na62exp} displays the setup upgraded from 2021 onwards. A $75\,\mathrm{GeV}$ momentum secondary beam, with a $6\,\%$ $K^+$ component, is produced by directing the $400\,\mathrm{GeV}$ proton beam from CERN SPS onto a beryllium target. The kaons are identified by a differential Cherenkov counter (KTAG). A silicon pixel detector based beam spectrometer (GTK) provides tracking of the beam particles. After the final collimator (COL) and the last station of GTK, the beam enters a vacuum tank, which includes the fiducial decay volume (FV). Downstream of it, the decay product $\pi^+$ is measured with a magnetic spectrometer (STRAW) for tracking, a ring-imaging Cherenkov counter (RICH) for particle identification, and two scintillator hodoscopes (CHOD) for timing and triggering. Additional particle identification is provided by a set of calorimeters consisting of a liquid krypton electromagnetic calorimeter (LKr), 2 hadronic iron/scintillator-strip sampling calorimeters (MUV1,2), and an array of scintillator tiles (MUV3) behind an iron wall. To reject any additional activity, comprehensive veto systems are employed. Upstream of the FV, scintillator-based veto detectors VC, CHANTI, and ANTI0 are designed to reject events with $K^+$ decays upstream of the fiducial volume, interactions between beam particles and materials, and beam halo muons, respectively. A photon veto system is composed of 12 large-angle veto (LAV) stations, an intermediate-ring calorimeter (IRC), and a small-angle calorimeter (SAC), and the LKr, covering angles from $0$ to $50\,\mathrm{mrad}$. Other veto detectors, including MUV0 for multi-track event suppression and HASC for photon conversion veto, are not visible in the side view of Fig. \ref{fig:na62exp}. 

In 2021--2022 data-taking, NA62 was operating under its maximum design intensity. Studies showed that the performance of the experiment saturates at high intensity~\cite{NA62SPSC24}. From August 2023, the beam intensity is reduced to $75\,\%$ of the maximum intensity, which provides the optimal performance according to the studies. Since 2023, the radiator gas of KTAG was changed from $\mathrm{N_2}$ to $\mathrm{H_2}$, lowering the material budget from $3.9\,\%\,X_0$ to $0.7\,\%\,X_0$~\cite{CedarH}. 

The KTAG, GTK, RICH, and CHOD provide excellent timing resolution at $\mathcal{O}(100\,\mathrm{ps})$. The combined RICH and calorimetric particle identification (PID) suppresses $\mu^+\Rightarrow\pi^+$ misidentification at $\mathcal{O}(10^7)$. The comprehensive photon veto system rejects $\pi^0$ at $\mathcal{O}(10^8)$. The excellent performance of the NA62 detector provides the sensitivity to the ultra-rare $K^+\to\pi^+\nu\bar\nu$ signal. 

\section{Event selections}

The analysis uses $K^+\to\pi^+\pi^0$ as the normalization channel. The normalization sample is selected from a dedicated trigger line (NORM) with a downscaling factor $D_\mathrm{NORM}=400$ and requires signatures consistent with a single $\pi^+$ track with in-time signal in KTAG. Offline the downstream $\pi^+$ track is matched to an upstream $K^+$ candidate in space and time. A vertex is defined as the mid-point of the Closest Distance of Approach (CDA) segment between the two extrapolated tracks. Any upstream activity is rejected using the upstream veto detectors described above, as well as information from GTK. A Boosted Decision Tree (BDT) classifier based on the spatial information of the two tracks is used to further reject events with the $\pi^+$ coming upstream of the FV. Finally, the full PID based on RICH and calorimeters is applied. 

The signal is selected from the signal trigger line (PNN) with no downscaling and applies trigger level photon and multi-track vetos in addition to the NORM trigger. In offline selections, in addition to the normalization selection, full photon veto and multi-track veto are applied. 

In the 2023--2024 dataset analysis, a transformer-based 4D GTK beam-tracking algorithm is developed, reducing the probability that the true $K^+$ is not reconstructed in the GTK from $6\,\%$ to $4\,\%$ with no significant loss of efficiency. Furthermore, a Convolutional Neural Network (CNN) based calorimetric PID is developed using information from the LKr and MUV1,2, significantly improving the performance against $\mu^+\Rightarrow\pi^+$ misidentification, especially when a photon cluster overlaps with the $\mu^+$ track in LKr. 

In PNN trigger line, the LAV is used to apply a trigger-level photon veto at large angles. Since 2023, this trigger condition was changed from vetoing the event with any signal in LAV stations 2--11 within a $6\,\mathrm{ns}$ time window to vetoing events within $4\,\mathrm{ns}$ time window and only using LAV stations downstream of the vertex between the $\pi^+$ track and the nominal beam direction. This change keeps $K^+\to\pi^+\pi^0$ background under control, but significantly increases the number of events with the $\pi^+$ coming from beam-material interactions upstream of the FV. To suppress these upstream background events, an offline veto using the first station of LAV within $2\,\mathrm{ns}$ time window is applied. This offline veto provides a much stronger suppression of these events than the previous trigger-level veto. 

\section{Signal sensitivity} 

The analysis exploits the kinematic variable squared missing mass, $m^2_\mathrm{miss}=(P_K-P_\pi)^2$, where $P_K$ and $P_\pi$ are the 4-momenta of the $K^+$ and $\pi^+$ candidates. In the $m^2_\mathrm{miss}$ distribution, signal regions are defined to suppress the main kaon decay modes at $\mathcal{O}(10^4)$~\cite{Pnn2017paper,PnnRun1Paper}. The analysis is performed in six categories of $5\,\mathrm{GeV}/c$ wide bins of $\pi^+$ momentum between $15$--$45\,\mathrm{GeV}/c$. For a specific $\pi^+$ momentum bin $p_i$, the observed number of events in signal region (SR) is given by
\begin{equation}
    N^{obs}_\mathrm{SR}(p_i) = N_{\pi\nu\bar\nu}(p_i) + N_{bg}(p_i) = \mathcal{B}(K^+\to\pi^+\nu\bar\nu)/\mathcal{B}_\mathrm{SES}(p_i) + N_\mathrm{bg}(p_i),
\label{eq:Br}
\end{equation}
where $N_\mathrm{bg}(p_i)$ is the expected number of background events in SR in bin $p_i$, and $\mathcal{B}_\mathrm{SES}$ is the Single Event Sensitivity:
\begin{equation}
    \mathcal{B}_\mathrm{SES}(p_i) = \frac{\mathcal{B}(K^+\to\pi^+\pi^0)A_{\pi\pi}(p_i)}{D_\mathrm{NORM}\cdot N_{\pi\pi}(p_i)\varepsilon_\mathrm{RV}\varepsilon_\mathrm{trig}(p_i)A_{\pi\nu\bar\nu}(p_i)}. 
\label{eq:ses}
\end{equation}
Here $N_{\pi\pi}$ is the selected number of normalization events, $\mathcal{B}(K^+\to\pi^+\pi^0)=(20.67\pm0.08)\,\%$ is the branching ratio of the normalization channel~\cite{PDG}, $A_{\pi\pi}$ and $A_{\pi\nu\bar\nu}$ are the acceptances of the normalization and signal selections, $\varepsilon_\mathrm{trig}$ is the trigger efficiency ratio of the PNN and NORM trigger lines, and $\varepsilon_\mathrm{RV}$ is the efficiency of the additional selections applied for signal channel with respect to the normalization channel. The inefficiency of the additional selections on signal events comes from unrelated activity which only depends on the beam intensity. 

A detailed evaluation procedure of these factors is presented in Ref.~\cite{Pnn2122paper}. All of the factors, summed or averaged over momentum bins, are shown in Table \ref{tab:results} left. The reduced intensity mentioned in Sec. \ref{sec:na62exp} results in a $14\,\%$ higher $\varepsilon_\mathrm{RV}$ and therefore better signal sensitivity.

\begin{table}[t]
    \caption[Summary tables]{Summary of the preliminary results of the 2023--2024 dataset analysis. Left: Summary of single event sensitivity inputs. $N_K=D_\mathrm{NORM}N_{\pi\pi}/\mathcal{B}(K^+\to\pi^+\pi^0)A_{\pi\pi}$ is the number of effective $K^+$ decays in this sample. $N_{\pi\nu\bar{\nu}}^\mathrm{SM}=\mathcal{B}^\mathrm{SM}(K^+\to\pi^+\nu\bar\nu)/\mathcal{B}_\mathrm{SES}$ is the expected number of SM $K^+\to\pi^+\nu\bar\nu$ events in SR. Right: the expected number of background events in SR, summed over the six $\pi^+$ momentum bins.}
    \label{tab:results}
    \centering
    \begin{tabular}{|c|c|} 
    \hline
    Factor & Value \\
    \hline
    \hline 
    $N_{\pi\pi}$ & $3.927\times10^{8}$ \\ 
    $A_{\pi\pi}$ & $(12.971\pm0.009)\%$ \\
    $N_{K}$ & $(5.93\pm0.02)\times10^{12}$ \\ 
    \hline
    $A_{\pi\nu\bar\nu}$ & $(7.36\pm0.33)\%$ \\
    $\varepsilon_\mathrm{trig}$ & $(86.4\pm1.2)\%$ \\
    $\varepsilon_\mathrm{RV}$ & $(72.3\pm0.7)\%$ \\ 
    \hline
    $\mathcal{B}_\mathrm{SES}$ & $(3.67\pm0.18)\times10^{-12}$ \\  
    $N_{\pi\nu\bar{\nu}}^\mathrm{SM}$ & $22.9\pm1.1$ \\
    \hline  
    \end{tabular} \hspace{3mm}
    \begin{tabular}{|l|l|} 
    \hline
    Background & Events \\
    \hline
    \hline 
    $K^{+}\to\pi^{+}\pi^{0}(\gamma)$ & ~$1.19\pm0.10$  \\ 
    $K^{+}\to\mu^{+}\nu(\gamma)$ & ~$1.39\pm0.29$  \\ 
    $K^{+}\to\pi^{+}\pi^{+}\pi^{-}$ & ~$0.25\pm0.05$ \\
    $K^{+}\to\pi^{+}\pi^{-}e^{+}\nu$ & ~$1.59^{+0.51}_{-0.43}$ \\
    $K^{+}\to\pi^{+}\gamma\gamma$ & ~$0.04\pm0.04$ \\
    $K^{+}\to\pi^{0}\ell^{+}\nu$ & ~$<0.001$  \\ 
    Upstream & ~$7.4^{+2.8}_{-2.2}$ \\
    \hline
    Total & \hspace{-2pt}$11.9^{+2.9}_{-2.3}$ \\
    \hline
    \end{tabular}
\end{table}

\section{Background evaluation}

Backgrounds from $K^+\to\pi^+\pi^0$, $K^+\to\mu^+\nu$, and $K^+\to\pi^+\pi^+\pi^-$ decays in the FV, which are highly suppressed kinematically, are evaluated by multiplying the number of events passing signal selection in their corresponding background regions by a kinematic tail fraction. The tail fraction is evaluated using a background control sample, by reconstructing $\pi^0$ and matching the vertices, by requiring $\mu^+$ PID, or using $K^+\to\pi^+\pi^+\pi^-$ Mote Carlo (MC) simulation, respectively. Other $K^+$ decays in the FV are evaluated with MC simulations, following the same procedure as computing $N_{\pi\nu\bar{\nu}}^\mathrm{SM}$ in Table \ref{tab:results}. The details of the evaluation procedures for background from $K^+$ decays in the FV can be found in Refs.~\cite{Pnn2122paper,Pnn2017paper}. 

The upstream background, in which the $\pi^+$ is coming from upstream of the FV, is evaluated using the CDA distribution. An Upstream Reference Sample (URS) is developed by removing the $K/\pi$ matching criteria  from the signal selection and inverting the requirement on CDA to $\mathrm{CDA}>4\,\mathrm{mm}$. The number of upstream events in the signal region is extrapolated from the URS with a template fit using two templates developed by enhancing the two different mechanisms: beam-material \emph{interaction}, and upstream decay with the $\pi^+$ matching to an \emph{accidental} beam particle. The effect of the $K/\pi$ matching criteria requirement is recovered by applying a mis-matching probability factor, $P^\mathrm{match}$, as described in Ref.~\cite{Pnn2122paper}. A correction factor for this method is evaluated to account for the different behavior of the upstream background in the signal region and the URS in $K/\pi$ matching. 

\begin{figure}
    \centering
    \includegraphics[width=0.5\linewidth]{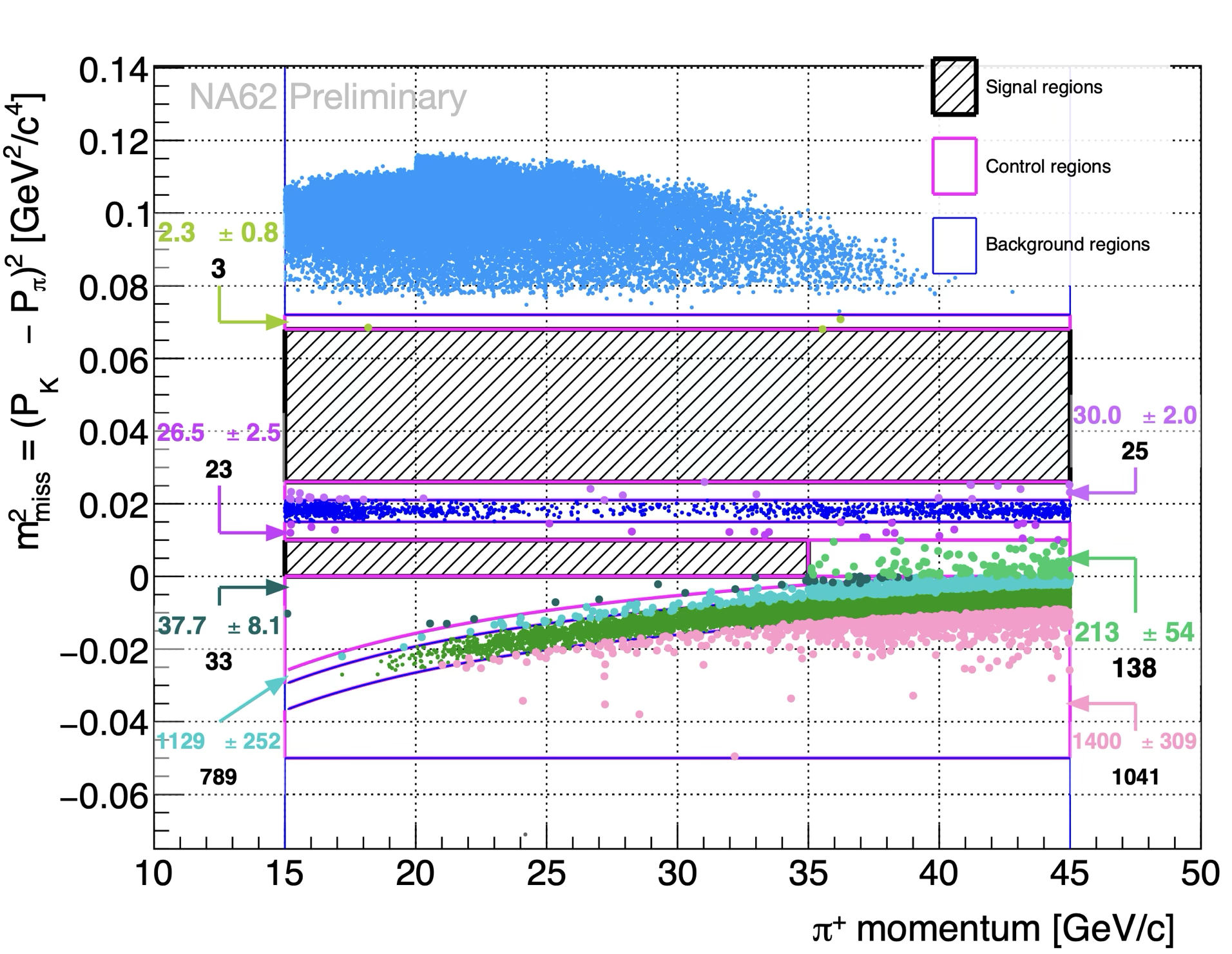}
    \includegraphics[width=0.39\linewidth]{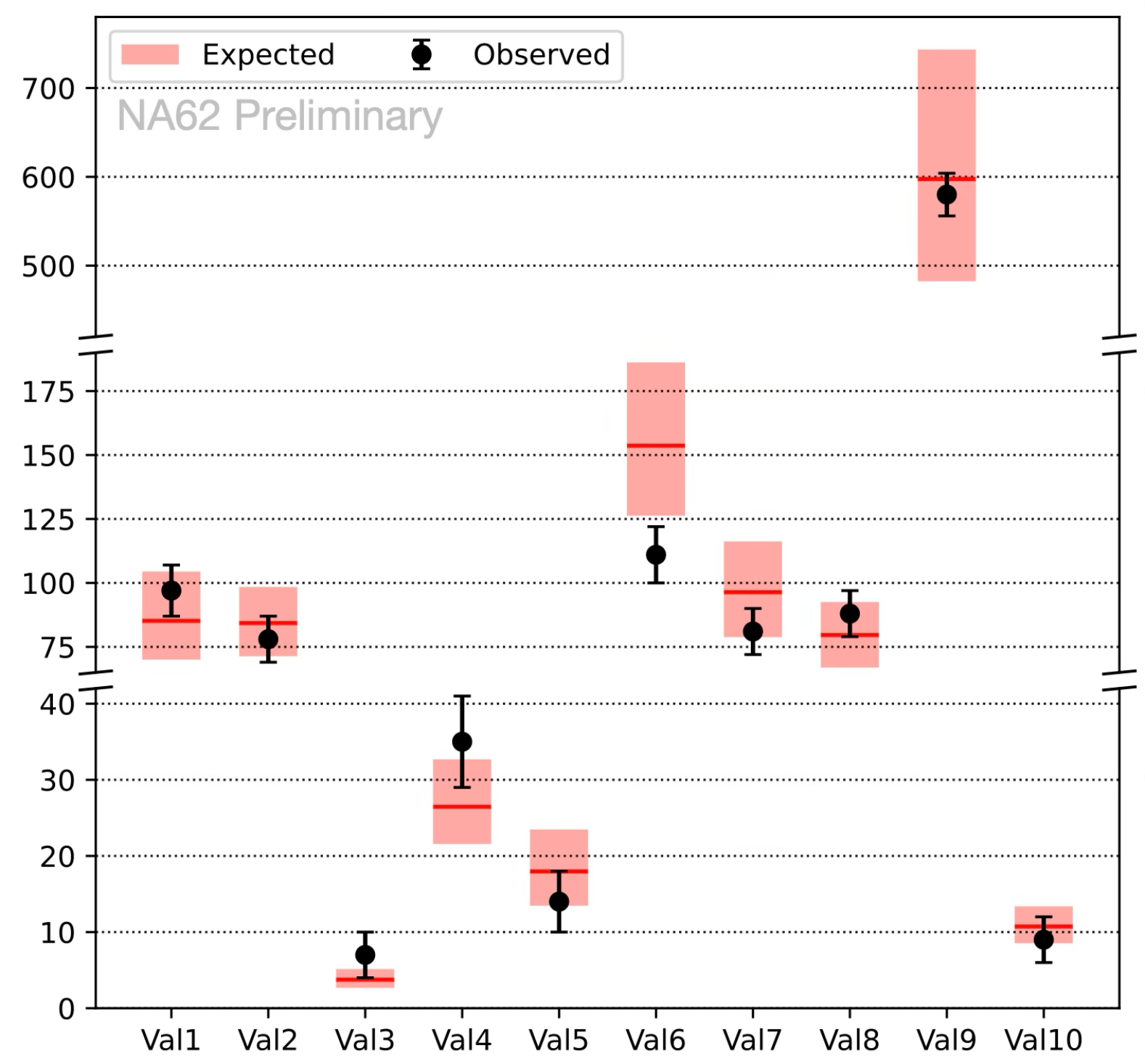}
    \caption[Background validation]{Background validation of the 2023--2024 dataset analysis. Left: control regions (defined in Ref.~\cite{Pnn2122paper}) with expectation (colored numbers) and observation (black numbers). Signal regions are blinded. Overall $p$-value $=0.65$. Right: upstream validation samples, developed by inverting different upstream veto conditions to enhance different mechanisms~\cite{Pnn2122paper}. Overall $p$-value $=0.79$.}
    \label{fig:bgval}
\end{figure}

A summary of the expected background is presented in Table \ref{tab:results} right. The validation of background evaluations are shown in Fig. \ref{fig:bgval}.

\section{Results}

\begin{figure}
    \centering
    \includegraphics[width=0.43\linewidth]{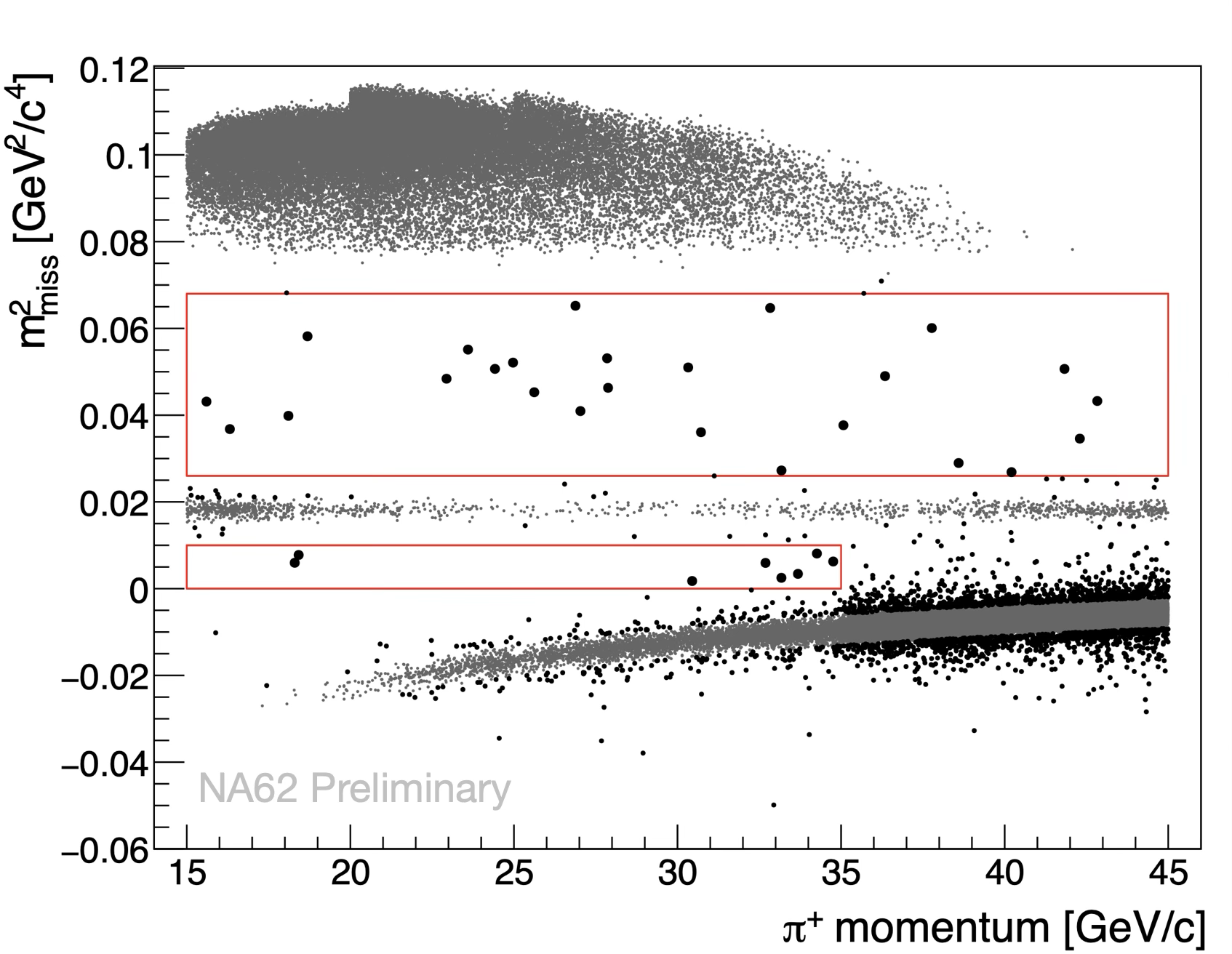}
    \includegraphics[width=0.5\linewidth]{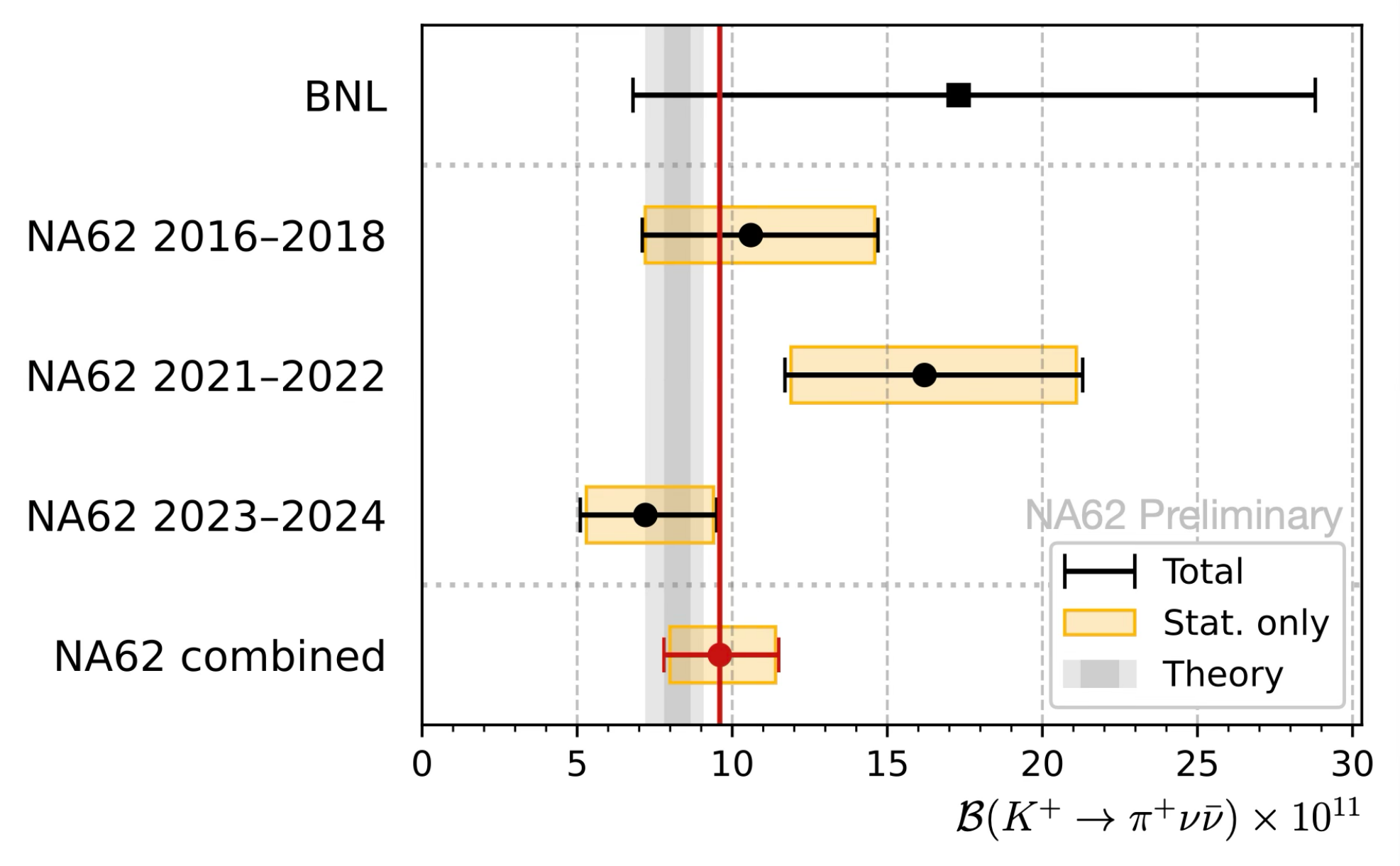}
    \caption[Results]{Left: The observed events in the 2023--2024 dataset satisfying the signal selection criteria. Right: summary of $K^{+}\rightarrow\pi^{+}\nu\bar{\nu}$ branching ratio measurements, compared to an envelope of the recent SM predictions~\cite{Buras:2022wpw,KaonsAtCERN2023,Allwicher:2024ncl}.}
    \label{fig:res}
\end{figure}

After unmasking the signal regions, $33$ candidate events are observed in the 2023--2024 dataset, shown in Fig. \ref{fig:res} left. By fitting Eq. (\ref{eq:Br}) to the $6$ categories, the branching ratio is measured to be
\begin{equation}
    \mathcal{B}_{2023-2024}(K^+\to\pi^+\nu\bar\nu) = (7.2^{+2.2}_{-1.9}|_\mathrm{stat}\ ^{+0.9}_{-0.9}|_\mathrm{syst})\times10^{-11} = (7.2^{+2.3}_{-2.1})\times10^{-11}.
\label{eq:Br2324}
\end{equation}
By combining these 6 categories with the 15 categories from the 2016--2022 data~\cite{Pnn2122paper}, a total of $84$ candidate events are observed in the 2016--2024 dataset with $30_{-3}^{+4}$ background events expected. By fitting the $21$ categories, the measured branching ratio gives
\begin{equation}
    \mathcal{B}_{2016-2024}(K^+\to\pi^+\nu\bar\nu) = (9.6^{+1.8}_{-1.6}|_\mathrm{stat}\ ^{+0.8}_{-0.6}|_\mathrm{syst})\times10^{-11} = (9.6^{+1.9}_{-1.8})\times10^{-11}.
\label{eq:Br1624}
\end{equation}

The branching ratios measured by the NA62 experiment with different datasets as well as the combined results from 2016--2024 dataset are shown in Fig. \ref{fig:res} right.  

\section{Conclusion}

The NA62 2023--2024 dataset doubled the effective sample size of the previous dataset while significantly reducing the background in proportion, bringing a major boost in sensitivity. The combined NA62 2016--2024 dataset provides a measured branching ratio $\mathcal{B}(K^+\to\pi^+\nu\bar\nu) = (9.6^{+1.9}_{-1.8})\times10^{-11}$, achieving a precision of $20\,\%$, excluding the background-only hypothesis with a significance exceeding $6\sigma$. The NA62 results are self-consistent and compatible with the results from the BNL E787 and E949 experiments~\cite{BNL-E949:2009dza}, and are in good agreement with SM predictions. The NA62 experiment continues to take data until 2026. Based on the number of days of data-taking, the whole dataset is extrapolated to increase by about $50\,\%$.

\section*{References}


\end{document}